\documentstyle[11pt,newpasp,twoside,epsf]{article}

\begin{document}
\title{Starburst-driven galactic superwinds}
\author{Dave Strickland}
\affil{Department of Physics \& Astronomy, The Johns Hopkins University
	3400 North Charles Street, Baltimore, MD 21218, USA.}

\begin{abstract}
I provide an observational review of the properties of starburst-driven
galactic superwinds, focusing mainly on recent results pertinent to
the transport of metals and energy into the IGM.
Absorption-line studies are providing rich kinematic
information on both neutral and ionized gas in superwinds, with observed mass
flow rates similar to the star formation rate and outflow velocities
comparable to or greater than the escape velocity. 
{\it FUSE} observations of the O{\sc vi} doublet
provide previously unattainable information regarding outflow
velocities and radiative cooling rates in hot gas at $T \sim 3\times 10^{5}$
K. Emission from gas at temperatures of $10^{4}$ K and
$\sim 5 \times 10^{6}$ K is now being studied with unprecedented
spatial resolution using {\it HST} and {\it Chandra}, tracing the
complex interaction of the still-invisible wind of SN-ejecta with the
ambient ISM entrained into these outflows.
I discuss the implications of these observations for our
understanding of starburst-driven outflows.
\end{abstract}

%-----------------------------------------------------------------
\section{Introduction}
\begin{figure}[!t]
\plottwo{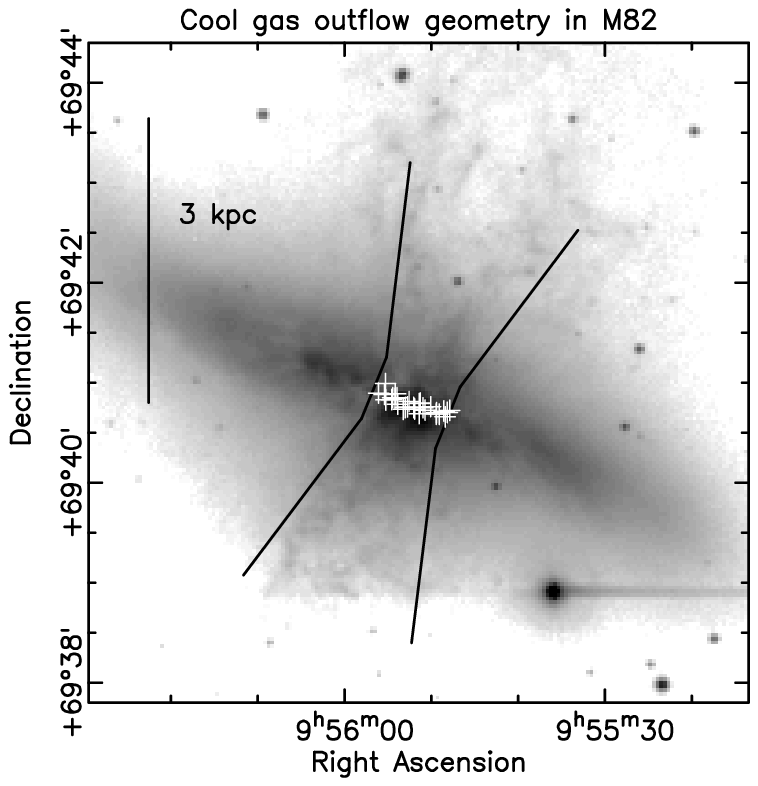}{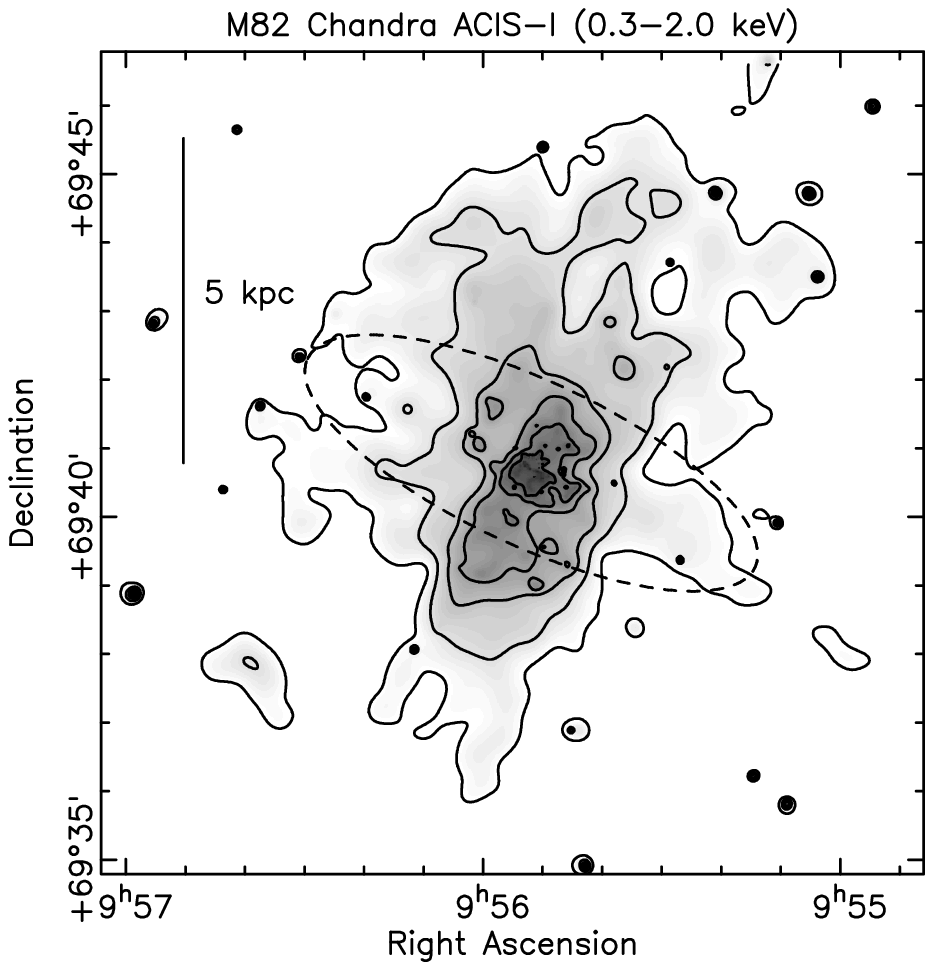}
\caption{(a) Narrowband H$\alpha$+continuum image of the nearby (D=3.63 Mpc)
	edge-on starburst galaxy M82, which shows filaments of $10^{4}$ K gas
	flowing out of the galaxy at $\sim 600$ km/s along the minor
	axis. The white crosses mark the positions of $\sim 50$ young SNRs 
	detected in radio observations of the central $800 \times 100$ pc 
	starburst region. 
	(b) A soft X-ray {\it Chandra} ACIS-I
	image of M82, showing
	gas with characteristic temperature of a few million
	degrees.}
\label{fig:m82_outflow}
\end{figure}

Observations of edge-on starburst galaxies show weakly collimated
10 kpc-outflows of gas (Fig.~\ref{fig:m82_outflow}), 
with outflow velocities of several hundred kilometers per second
(McCarthy, Heckman, \& van Breugel 1987; 
Heckman, Armus, \& Miley 1990; Armus, Heckman, \& Miley 1990).
Tracers of warm ionized gas such as H$\alpha$~emission show
filaments and arcs of emission extending outward from the nuclear regions
of the host galaxy galaxy, which outline the surfaces
of bipolar outflow cones of opening angle $\theta \sim 60 \deg$.
The primary observational probes of these outflows have historically been
been optical emission lines
(Armus et al. 1990), 
and X-ray emission (Dahlem, Weaver, \& Heckman 1998),
 although all phases of the ISM
have been detected (Dahlem 1997).
 The X-ray emission correlates well spatially
with the H$\alpha$~emission (see Fig.~\ref{fig:m82_outflow}), 
although in many cases the
X-ray observations trace these outflows out to larger
galactocentric radii ($\la 20$ kpc,
Read, Ponman, \& Strickland 1997) than the H$\alpha$ observations.

These outflows result from the  energy returned
to the ISM by the recently-formed massive stars in the starburst.
Core collapse supernovae and massive star stellar winds, from 
$\sim 10^{6}$ O \& B stars in galaxies like M82 or NGC 253, return large
amounts of kinetic energy along with metal-enriched ejecta to the
ISM. Radio observations of local starbursts reveal large numbers of
young SNRs within the starburst region
(Kronberg, Biermann, \& Schwab 1981; Muxlow et al. 1994).
%These radio sources are known to be young supernova remnants, as
%many of them are resolved and show clear
%shell-like morphologies (McDonald et al. 2001), 
%and they obey the LMC SNR radio
%surface brightness vs. radius relationship.
Age estimates for the starburst stellar populations 
($\sim 10$ Myr for M82, 20 -- 30 Myr 
in NGC 253 [Satyapal et al. 1997; Engelbracht et al. 1998]) 
agree well with the dynamical ages
 of the outflows, $\tau_{\rm dyn} \sim 10$ 
kpc/500 km/s $\sim 20$ Myr.
The kinetic energy of the individual remnants and wind-blown
bubbles is thermalized via shocks as SNRs overlap
and interact, creating a hot ($T \sim 10^{8}$ K), high pressure
($P/k \sim 10^{7}$ K cm$^{-3}$)
bubble of metal-enriched gas in the starburst region (Chevalier \& Clegg 1985).
This ``superbubble'' expands preferentially along the
path of least resistance (i.e. lowest density), 
breaking out of the disk of the galaxy along the minor axis after a
few million years. The hot gas then
expands at higher velocity ($v \ga 1000$ km/s)
into the low density halo of the galaxy as a superwind,
dragging along clumps and clouds of cool dense entrained ISM
at lower velocity (see Suchkov et al. 1994).

Many excellent reviews of both observations and theory 
of starburst-driven superwinds already exist 
(Heckman, Lehnert, \&  Armus 1993; Heckman 1998). 
In this contribution I highlight recent results 
related to the issue of mass, metal and energy transport
by superwinds out of galaxies and into the IGM.

%----------------------------------------------------------------------
\section{Starburst galaxies in the local universe}

The average starburst galaxy in the local universe (weighted by
FIR luminosity, a convenient estimator of
the star-formation rate (SFR) in all but the most metal-poor galaxies) 
has a FIR luminosity of $L_{\rm FIR} \sim L_{\star}$ 
(Soifer et al. 1987) and a SFR of 
a few $M_{\sun}$ yr$^{-1}$ (Heckman et al. 1993).
 In general these galaxies are late type spiral
galaxies with rotational velocities $v_{\rm rot} \sim 200$ km/s, 
local examples being NGC 253, NGC 3628 \& NGC 4945.
It is these ``typical'' starbursts
that show most clearly evidence for starburst-driven outflows 
(Armus et al. 1990), and I shall mainly concentrate on discussing
this class of galaxies.
In terms of overall significance, approximately $\sim 25$\% 
{\em of all high mass star-formation 
in the the local universe is in starburst
galaxies} (Heckman 1998), and it is likely that all starbursts
drive outflows. This is not a rare or exotic phenomenon. 

Although outflows from dwarf galaxies have captured much of the theoretical 
effort on outflows (e.g. Mac Low \& Ferrara 1999), under the assumption
that it is easier to drive outflows in low mass systems, it is
important to realize that this does not automatically mean 
that larger galaxies do not drive outflows.
%%Outflows from local dwarf starburst galaxies are observed, 
%%primarily based on H$\alpha$~imaging (e.g. 
%%Meurer et al. 1992; Marlowe et al. 1997; Martin 1997], but
%%they are generally smaller ($\la 3$ kpc in radius [e.g. Hunter \& Gallagher
%%1997]) and more sedate kinematically ($v_{{\rm H}\alpha} \la 100$ km/s
%%compared to $\sim 500$ km/s) than the typical superwinds
%%(Martin 1998).
I will not discuss outflows from local ultraluminous
IR galaxies (ULIRGs) which have SFRs
 up to $\sim 30$ times greater than the average starburst galaxy, 
or the starbursting systems seen at high
redshift (see Pettini, this volume), 
except to note that they  appear {\em at least}
 as powerful as winds in local starburst galaxies.

%------------------------------------------------------------------------
\section{The efficiency of supernova heating}
\label{sec:sn_heating}

The SN rate within a typical local starburst
galaxy is $\sim 0.05$ yr$^{-1}$ (Mattila \& Meikle 2001), 
which implies a total number of SNe exceeding
$\sim 2 \times 10^{6}$ over the lifetime of a starburst event.
How much of the kinetic energy from all these SNe is available 
to drive the wind?
This is of particular interest with respect to the possible role
SN-driven winds play in imparting additional heating to the IGM and ICM (see
various discussions in this volume).

It is useful to split the problem into two parts, firstly
radiative losses within the starburst region, and secondly
radiative losses within the larger-scale superwind.
The second of these can be assessed relatively
straightforwardly with observations of local superwinds, and will
be addressed in \S~\ref{sec:uv}
The radiative losses young SNRs suffer within the starburst region
are extremely difficult to determine observationally,
as these regions are heavily obscured. Consequently arguments about
radiative energy losses are based purely on theory and
a wide range of opinions exist. I shall present 
the situation as I see it, and refer the reader to the contribution
by Recchi (this volume) for a different point of view.

A single isolated SNR, evolving in a uniform medium of number density
$\sim 1$ cm$^{-3}$ will lose a fraction $f \sim 90$\% of 
its initial kinetic energy
to radiation over $\sim 4\times 10^{5}$ years (Thornton et al. 1998).
Adopting the terminology of Chevalier \& Clegg (1985),
this corresponds to a thermalization efficiency
of $\eta_{\rm therm} = 1-f \sim 10$\%. 
Cooling depends sensitively on the local
density as $L = \int n^{2} \Lambda dV$.
Many authors assume that because bursts of star formation occur 
in regions with large amounts of dense gas, virtually all 
the energy from SNe is lost due to radiation (e.g. Steinmetz 1999). 
This ignores the multiphase nature of the ISM where the filling
factor of dense gas is low, and that the
phase structure is determined by the 
local SN rate (Rosen \& Bregman 1995).

In a starburst region such as at the center of M82 or NGC 253
the SN rate per unit volume is a few $\times 10^{-9}$ yr$^{-1}$ pc$^{-3}$,
about 5 orders of magnitude higher than the SN rate/volume in the disk 
of the MW ($\sim 4 \times 10^{-14}$ yr$^{-1}$ pc$^{-3}$, 
Slavin \& Cox [1993]).
%Within individual super star clusters the SN
%rate/volume can be orders of magnitude higher than the volume-averaged mean
%quoted above.
The average individual SNR or 
wind blown bubble in a starburst 
does not exist long enough to radiate away 90\%
of its energy before it runs into another remnant or pre-blown
low density cavity. Once in a low density medium radiative losses
cease to be significant (see Mac Low \& McCray 1988).
As a consequence the thermalization
efficiency in starbursts {\em must} 
be considerably higher than the 10\%
value applicable to ``normal'' star-forming disks. 
Numerical simulations investigating thermalization efficiency as a function
of SN rate/volume support this argument (Strickland, in preparation).
Some SNe may occur  molecular cores, and suffer significant
radiative losses, but on average SNe in the starburst do not lose
a large fraction of their energy. Thermalization efficiencies
 $\eta_{\rm therm} \ga 50$\% are quite possible.

In principle, observationally measuring the temperature of the very hot
tenuous gas in the starburst region (i.e. the thermalized 
SN ejecta) can directly provide the thermalization efficiency.
Using the rates of mass and energy input from the Starburst99
models (Leitherer et al. 1999), $T_{\rm gas} =  1.2 \times
10^{8} \eta_{\rm therm} \beta^{-1}$ K, where $\beta \ge 1$ 
is a measure of mass-loading (Suchkov et al. 1996).
The faint X-ray emission from this hot gas can, in principle,
be detected in  nearby starburst galaxies. Unfortunately starburst
regions are also host to large numbers of
of X-ray binaries, and possible low luminosity AGN,
making this measurement extremely difficult.
The first  believable detection of this very hot gas
 uses {\it Chandra}'s high spatial resolution to resolve out the
X-ray binaries. Griffiths et al. (2000) claim to detect diffuse emission from 
a $T \sim 5 \times 10^{7}$ K gas within M82's starburst region, 
which if confirmed implies $\eta_{\rm therm} \sim 40 \beta^{+1}$\%.

%------------------------------------------------------------------------

\section{Warm neutral gas in superwinds}
\label{sec:wnm}

\begin{figure}[!t]
\plotone{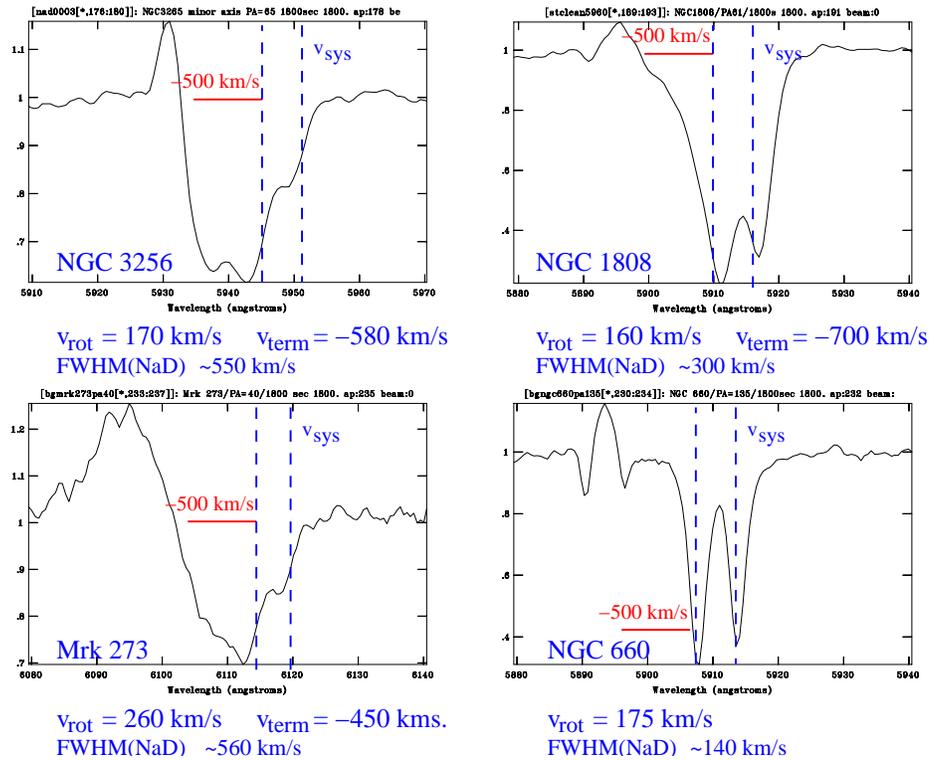}
\caption{A representative sample of Na D absorption line
profiles for four starburst galaxies, adapted from the larger
sample shown in Heckman et al. (2000). 
The dashed vertical lines show the expected centroids of
the Na D doublet at the systemic velocity of the galaxy.
Horizontal bars represent a blue
shift of 500 km/s. Note the strongly blue-shifted Na D absorption in 
NGC 3526, NGC 1808 and Mrk 273 due to the superwinds in these galaxies.}
\label{fig:nad}
\end{figure}

Many local starburst galaxies show blue-shifted
Na D absorption line profiles (Phillips 1993; Heckman et al. 2000).
The sodium D (Na{\sc i} $\lambda\lambda$ 5890, 5896) 
lines probe warm neutral gas, at a temperature
of a few thousand degrees.
In Heckman et al's sample of IR luminous starburst galaxies 19 out
of a sample of 33 showed blue shifted absorption,  typically extending
out to terminal velocities between $v_{\rm term} = -200$ to $-700$ km/s 
(Fig.~\ref{fig:nad})
in galaxies with rotation velocities between 140 and 330 km/s.
Absorption is seen over a wide range in velocity, from the systemic
velocity of the galaxy out to $v_{\rm term}$, suggesting gas with
multiple velocity components in the outflow.
Those galaxies that show blue-shifted absorption tend to be
more face-on than those that do not show absorption.

This is consistent with a model where
the blue-shifted absorption features arise in cool ambient gas entrained
into a weakly collimated outflow along the minor axis of the galaxy.
The gas initially has low velocity, giving rise to absorption
near the systemic velocity of the galaxy, but is accelerated to higher
velocity by the ram pressure of the SN-ejecta wind.

This cool gas dominates the total mass in the outflow. For a typical
starburst in this sample ($L_{\rm FIR} \sim 2\times 10^{11} L_{\sun}$)
the mass of warm neutral gas is $M_{\rm NaD} \sim 5 
\times 10^{8} M_{\rm sun}$. The mass flow rate in
this component significantly exceeds the mass injection rate 
due to SNe and stellar winds and is comparable to the gas consumption rate
due to star formation ($\dot M_{\rm NaD} \sim 3$ -- 
$10 \times \dot M_{\rm SN} \sim \dot M_{\rm SF}$). Although
not highly metal-enriched, this component is significant in terms of total
mass of metals transported out of the disk of the galaxy.
%The amount of 
%energy required to accelerate this gas to the observed velocities is only
%$\sim 1$ -- $10$\% of the total energy released by SNe (assuming
%$\eta_{\rm therm} = 1$).

Does this gas escape the galaxy and pollute the IGM? The observed terminal
velocities are typically several times the rotational velocity of the galaxy,
comparable to or greater than the galactic escape velocity assuming
$v_{\rm esc} \sim 3\times v_{\rm rot}$. It should be stressed that
even if $v < v_{\rm esc}$, this does not automatically 
imply that the gas is retained by the galaxy.
The motion of this cool gas is not simply ballistic, as the
clouds are carried along by the wind. 
The long term fate of this gas is unknown, and depends more on
on hydrodynamic forces (wind ram pressure, retardation by halo gas)
than the gravitational potential of the galaxy.

\section{Warm ionized gas}
\label{sec:wim}
A large literature exists using optical emission lines to study
warm ionized gas at $T \sim 10^{4}$ K in superwinds (see Heckman et al. 
1993 and references therein). Given this, I will only briefly
mention some of the important wind diagnostics provided by
these studies, before discussing what I believe to be an
important observation affecting numerical estimates of
mass loss from superwinds.
 
Balmer lines, primarily H$\alpha$~emission,
provide kinematic information along with
morphological information regarding the structure of the outflow.
Spatially resolved kinematic studies (McKeith et al. 1995; Shopbell
 \& Bland-Hawthorn 1998; Cecil et al. 2001) 
provide information
on the entrainment and acceleration of cool gas.
The [S{\sc ii}] doublet 
($\lambda\lambda$ 6717, 6731) can be used as a density diagnostic.
This has been used to derive densities and pressures in the warm 
clouds in superwinds (McCarthy et al. 1987; Armus et al. 1990). 
Line ratios also provide ionization source diagnostics. The gas near the
starburst region is primarily photoionized by the intense UV radiation
from the massive stars, but at larger
distances the H$\alpha$~emitting gas shows line ratios indicative
of shock heating (Martin 1997).
In the future we may hope to apply the detailed
shock diagnostics used in studies of local SNRs to superwinds.

One particularly noteworthy recent development is high resolution
studies with {\it HST} of the optical emission line filaments in NGC 3079
's superwind (Cecil et al. 2001) and M82's superwind (Shopbell et al,
in preparation). At high resolution the filaments and arcs of 
warm gas break up into very small clumps or clouds. The
largest clumps or clouds  in NGC 3079 are $\sim 30$ pc in diameter,
although many are unresolved even by HST.  Most of the mass in a superwind
is in these cool, very compact clouds. Accurately treating the entrainment
and acceleration of such clouds by the wind requires that the model
resolves the wind/cloud interaction.
Klein, McKee, \& Colella (1994) argue that
this requires at least 100 cells across across a cloud diameter,
implying cell sizes of $\ll 1$ pc in simulations that must cover
2-or-3-dimensional volumes 10's of kpc on a side (preferably $\sim 200$ kpc
on a side, see Aguirre and Pettini in this volume).
No current simulations achieve this level of resolution, and 
therefore these
models may significantly underestimate mass loss
in superwinds.

\section{UV absorption probes of coronal gas}
\label{sec:uv}

The launch of NASA's Far Ultraviolet Spectral Explorer ({\it FUSE})
mission in 1999 has finally allowed absorption from the O{\sc vi}
$\lambda\lambda$ 1032, 1038 doublet to be detected in local starburst
galaxies. These lines probe collisionally ionized gas at $T \sim 3 \times
10^{5}$ K. These observations provide us with kinematic
information on gas $\sim 30$ times hotter than probed by 
optical emission lines.

A theoretical prediction of both analytical and numerical
models of superwinds is that hotter gas has higher outflow velocities
than the cool gas. The maximum velocities are achieved by the
very energetic SN-ejecta, while cooler denser ambient ISM
is swept up and accelerated to a terminal velocity dependent on
the column density of the clump or cloud (Chevalier \& Clegg 1985).
If true, hot gas might escape these galaxies even if 
outflow velocities in the cool phases are well below escape velocity.

If superwinds are to be stopped before reaching the IGM, it
is necessary to radiate away the majority of their energy. 
Observations place strong limits on the radiative
power loss in the X-ray band at $\sim 1$\% or so of the SN energy
injection rate $\dot E_{\rm SN}$, while optical emission lines account for a
few to $\sim 10$\% of $\dot E_{\rm SN}$. The only waveband 
where appreciable energy could be emerging that has not previously
been explored  is the far UV.

Blue-shifted absorption from O{\sc vi} is seen in a variety of 
of starbursts of different mass and star-formation intensity, 
from starbursting dwarf galaxies like NGC 4214 (Martin et al, in preparation) 
through to typical starbursts like NGC 3310.
Heckman et al. (2001) present a detailed case study of one the archetypal
starbursting dwarf galaxies, NGC 1705, which shows
a complicated optical  morphology suggesting that
hot gas is in the process of ``blowing-out'' of 
a $\sim 2$ kpc diameter H$\alpha$ bubble. The {\it FUSE} observations
reveal that the hot gas responsible
for the O{\sc vi} absorption has a higher outflow velocity than
the warm ionized medium, which in turn has a higher outflow velocity than
warm neutral gas ($v_{\rm OVI} = -77\pm{10}$ km/s, 
$v_{\rm WIM} = -53\pm{10}$ km/s, $v_{\rm WNM} = -32\pm{11}$ km/s).
These observations are inconsistent with the standard superbubble
model (Castor, Weaver, \& McCray 1975; Mac Low \& McCray 1988), but agree with
the predictions gas entrainment and acceleration as hot gas flows
out through holes in a fragmented superbubble shell to form
a superwind. FUV radiative losses in NGC 1705 appear minimal, 
only $\sim 5$\% of $\dot E_{\rm SN}$, so superwinds appear to be
inefficient radiators at any wavelength.

\section{Thermal X-ray emission from superwinds}
\label{sec:xray}

\begin{figure}[!t]
\plotone{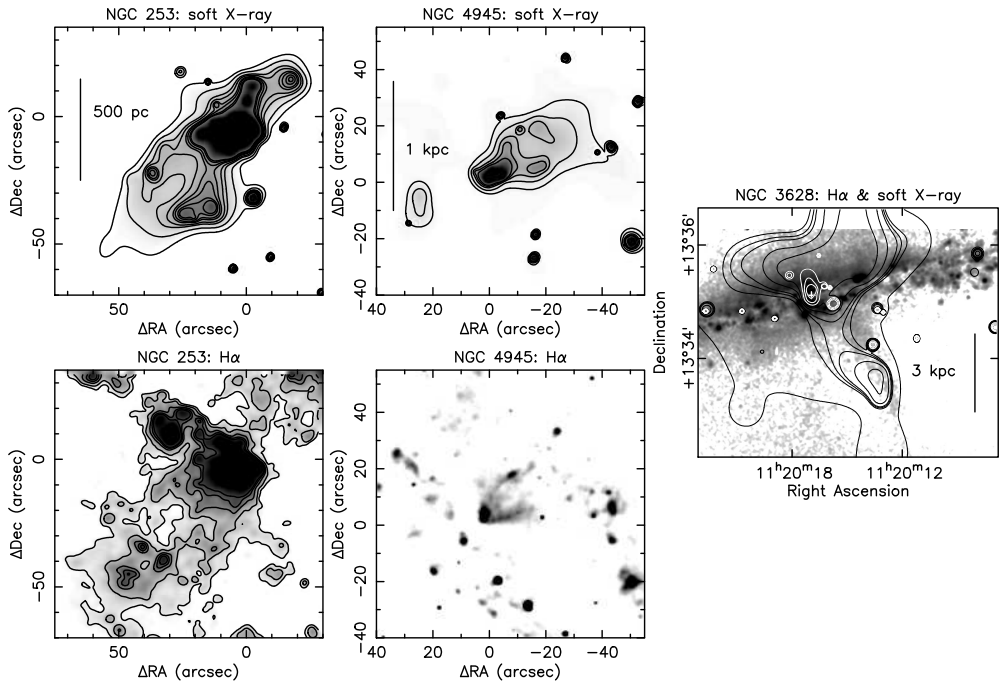}
\caption{Soft X-ray and H$\alpha$ emission in several
	edge-on starburst galaxies, showing the
	spatial similarities between the two phases.
	NGC 253 \& NGC 4945 have kpc-scale
	limb-brightened nuclear outflow cones
	(the opposite outflow cone is obscured in both cases) 
	with a close match between X-ray
	\& H$\alpha$~emission. In NGC 3628 a $5$ kpc-long
	H$\alpha$ arc on the eastern limb of the wind
	is matched by an offset X-ray filament.}
\label{fig:xray_halpha}
\end{figure}

The motivation for studying X-ray emission from superwinds has always
been the hope that the observed thermal X-ray emission provides a direct
probe of the hot, metal-enriched, gas that drives these outflows.
The 10-kpc-scale diffuse X-ray emission in superwinds seen by {\it Einstein}, 
{\it ROSAT} \& {\it ASCA} had characteristic temperatures of
a few million degrees. This is much cooler than the
$\sim 10^{8}$ K expected for raw SN-ejecta within the starburst region,
but as the wind expands cooling processes such as adiabatic expansion
 or mass-loading (Suchkov et al. 1996)
might reconcile the observed temperatures with an interpretation of the
X-ray emission coming from a volume-filling 
wind of SN-ejecta. The alternative model
is that X-ray emission comes from an interaction between
the hot, high velocity wind, and cooler denser ISM along the walls of the
outflow or in clouds embedded within the wind (Chevalier \& Clegg 1985;
Suchkov et al. 1994; Strickland \& Stevens 2000).
A fraction of the cool dense ISM  is shock-or-conductively 
heated to million degree temperatures.

Distinguishing between these two competing models has awaited high spatial
resolution X-ray imaging. If the X-ray emission comes from a volume-filling
wind, then the X-ray emission should smoothly fill the interior of the
cone or lobes outlined by H$\alpha$~emission. In the wind/ISM interaction
model the X-ray emission should be concentrated in regions with dense
cool gas, and it should appear as filamentary and 
limb-brightened as the H$\alpha$~emission. The spatial resolution and
sensitivity of X-ray instruments prior to {\it Chandra} was not
high enough to make an exact comparison between X-ray and H$\alpha$ emission
in even the nearest superwinds, although a general correlation between the
two has long been noted (Watson et al. 1984; McCarthy et al. 1987).

{\it Chandra}'s $\la 1\arcsec$ spatial resolution corresponds to $\sim 12$
pc at the distance of nearby superwinds -- the same physical scales as the
H$\alpha$~emitting clouds. We now have unambiguous 
evidence (Fig.~\ref{fig:xray_halpha}) for a 1-to-1 relationship between
the spatial distribution of the soft thermal X-ray emission and the
H$\alpha$ emission in the inner kpc of several superwinds (Strickland
et al. 2000; 2001 in preparation). Over the
larger 10 kpc scales of the winds the X-ray emission still appears 
filamentary and arc-like, and is associated with nearby filaments or
arcs of H$\alpha$ emission. Low-volume filling factor
gas dominates the X-ray emission from superwinds, and  the
the gas that actually drives the outflow remains invisible.

\section{Summary and conclusions}

Typical starburst galaxies ($L_{\rm FIR} \sim L_{\star}$) show high
velocity (200--700 km/s) multi-phase outflows.
The observed mass flow rates in the wind are comparable to the
gas consumption rate due to star formation ($\dot M_{\rm wind} 
\ga \dot M_{\star}$), and are dominated by relatively cool ambient
gas ($T \sim$ few thousand K) that has been swept-up and accelerated
by the ram pressure of the hotter wind of SN-ejecta. 

Outflow velocities are typically comparable or greater than estimates
of the galactic escape velocity, but caution should be
exercised in making claims about mass loss rates. 
The gas motions are not ballistic, making it  impossible to
give quantitative observational {\em mass loss} rates. Observations
of gas at much larger galactocentric radii ($\ga 100$ kpc) are
needed to directly observe mass loss. Nevertheless, the observed
{\em mass flow} rates are considerable, and does seem likely that 
some significant fraction of even the coolest phases may well escape
even moderately massive starburst galaxies.
Speaking as a practitioner of hydrodynamical simulations of superwinds, I am
not convinced that we know mass-loss rates theoretically. Existing
models have yet to be meaningfully tested
against observations. The lack of sub-parsec
numerical resolution in current simulations prejudices
the ability of these models to treat mass transport in winds.

{\it FUSE} observations of O{\sc vi} absorption provide vital information
of the kinematics and radiative losses of coronal 
gas at $T\sim 3 \times 10^{5}$ K.
In the dwarf starburst NGC 1705 the {\it FUSE} observations support
the theoretical prediction of superwind models that the hotter phases
in superwinds have higher outflow velocities than the cooler phases.
This suggests that the even hotter material holding the metals is more
likely to escape than the warm neutral and ionized gas.
Radiative energy losses within the wind appear minimal
compared to the energy injection rate from SNe,
even within the FUV 
and X-ray wave-bands.

With the sub-arcsecond spatial resolution provided by {\it Chandra}
it is now clear that the soft thermal X-ray emission seen in
superwinds is due to some form of interaction between the 
(still invisible) high
velocity hot wind and cooler denser ambient gas swept-up or
overrun by the flow. This is unfortunate in the sense that
X-ray observations do not provide a direct probe
of the energetic metal-enriched gas driving these winds.
Nevertheless the data from {\it Chandra} allow us to obtain more accurate
estimates of the physical properties of the X-ray emitting gas
than ever before, and provide deeper insight into the conditions
within these winds.

Starburst-driven winds are difficult objects to study,
due to the range of different gas phases involved and the
faintness of the emission. Nevertheless, 
very significant progress
is being made, most notably due to the new observational capabilities
provided by {\it FUSE} \& {\it Chandra}. The wealth of multi-wavelength
data will place extremely strong constraints upon numerical models
of superwinds. This is an exciting time to study
superwinds, and there is no prospect of an end to new discoveries 
about these fascinating and important objects.

\acknowledgments

It is a pleasure to thank Tim Heckman for countless enlightening
discussions over the years, Crystal Martin and Gerhardt Meurer
for providing a variety of spectra and images, and Francesca Matteucci
for organizing a stimulating workshop. DKS gratefully acknowledges the support 
from {\it Chandra} Postdoctoral Fellowship Award
Number PF0-10012, issued by the {\it Chandra} X-ray Observatory Center,
operated by the SAO on behalf of NASA.

\end{document}